\documentclass[twocolumn,floatfix,prb,amsmath,aps]{revtex4}
\usepackage{graphicx}
\usepackage{bm}
\usepackage{amssymb}
\begin{document}
\bibliographystyle{apsrev}
\def\nn{\nonumber}
\title{Topological chiral magnonic edge mode in a magnonic crystal} 
\date{\today}
\author{Ryuichi Shindou} 
\altaffiliation{present affiliation: International Center for 
Quantum Materials (ICQM), Peking University, No.5 Yiheyuan 
Road, Haidian District, Beijing, 100871, China.} 
\affiliation{Department of Physics, Tokyo Institute of Technology,
2-12-1 Ookayama, Meguro-ku, Tokyo, Japan} 
\author{Ryo Matsumoto} 
\affiliation{Department of Physics, Tokyo Institute of Technology,
2-12-1 Ookayama,  Meguro-ku, Tokyo, Japan} 
\author{Jun-ichiro Ohe}
\affiliation{Department of Physics, Toho University, 
2-2-1 Miyama, Funabashi, Chiba, Japan}
\author{Shuichi Murakami} 
\affiliation{Department of Physics, Tokyo Institute of Technology,
2-12-1 Ookayama, Meguro-ku, Tokyo, Japan}  
\begin{abstract}
Topological phases have been explored 
in various fields in physics such as 
spintronics, photonics, liquid helium, correlated 
electron system and cold-atomic system. This 
leads to the recent foundation of emerging materials 
such as topological band insulators, 
topological photonic crystals and 
topological superconductors/superfluid. In this paper,  
we propose a topological magnonic crystal which provides 
protected chiral edge modes for magnetostatic 
spin waves. Based on a linearized Landau-Lifshitz equation, 
we show that  a magnonic crystal with the dipolar 
interaction acquires spin-wave volume-mode band 
with non-zero Chern integer. We argue that 
such magnonic systems are accompanied 
by the same integer numbers of 
chiral spin-wave edge modes within a 
band gap for the volume-mode bands. 
In these edge modes, the 
spin wave propagates in a unidirectional  
manner without being scattered backward, 
which implements novel fault-tolerant spintronic 
devices.
\end{abstract}
\maketitle
\section{introduction}
Topological phases in condensed matters  
have been attracting much attention because 
of their fascinating physical properties. Discoveries of 
topological band insulators~\cite{TI1,TI2,TI3,TI4,TI5,TI6,TI7} 
open up emerging research 
paradigm on spin-orbit interaction physics. 
Relativistic spin-orbit interaction in the 
topological band insulator endows 
its Bloch electron bands with non-trivial global 
phase structures, which lead to novel surface metallic 
states.~\cite{TI5,TI8} 
Superconductor analogues of topological insulators 
have exotic edge modes~\cite{TS1,TS2}, which 
are composed only of Majorana fermion. Some aspects 
of these bound states are experimentally 
confirmed,~\cite{TS3,TS4} fostering 
much prospect of the realization of 
quantum computers.~\cite{TS5} 
Photonics analogue of quantum Hall phase 
with chiral edge modes are proposed 
theoretically~\cite{Hal1,Hal2,cZhang} 
and are subsequently designed in actual photonics 
crystals.~\cite{Wang} Unidirectional 
propagations of electromagnetic wave along these 
edge modes were experimentally observed, 
which provides these metamaterials with 
unique photonic functionality.  

In this paper, we theoretically propose a spin-wave  
analogue of topological phases, which has 
topologically-protected chiral edge mode for the 
spin wave propagation. Spin wave is a collective propagation 
of precessional motions of magnetic moments   
in magnets. Depending on its wavelength,  
spin waves are classified into two 
categories. One is exchange  
spin-wave with the shorter 
wavelength, whose motion is driven by the 
quantum-mechanical exchange 
interactions (`exchange-dominated' region). 
The other is magnetostatic spin 
wave with the longer wavelength,~\cite{Damon1,Damon2}  
whose propagation is caused 
by the long-range dipolar interaction  
(`dipolar' region).  
Magnonics research investigates how these 
spin waves propagate in the sub-micrometer  
length scale and sub-nanosecond time 
scale.~\cite{Kruglyak,Serga,Lenk} 
Like in other solid-state technologies 
such as photonics and plasmonics, 
the main application direction 
is to explore ability of the spin wave to carry 
and process information. 
Especially, the propagation of spin waves  
in periodically modulated magnetic materials 
dubbed as 
magnonic crystals~\cite{Kalinikos,MC1,MC2,MC3,Lenk,Gulyaev,
Singh,ccWang,Adeyeye,Matsumoto1,Matsumoto2} 
are of one of its central concern. 
Owing to the periodic structuring, the 
spin-wave volume-mode spectrum in magnonic crystal acquires 
allowed frequency bands of spin wave states 
and forbidden-frequency bands 
(band gap).~\cite{Kalinikos,MC1,MC2,MC3,Lenk,Gulyaev,
Singh,ccWang,Adeyeye}

\begin{figure}
\begin{center}
\includegraphics[scale=0.38]{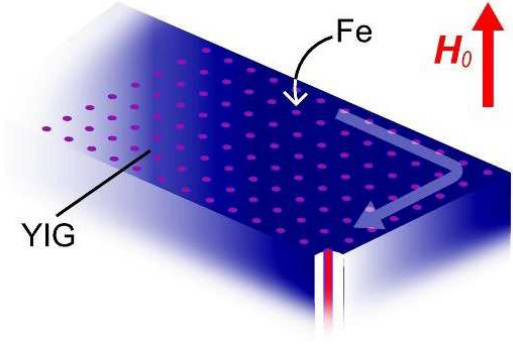}
\caption{Magnonic crystal with chiral edge modes.     
Periodic array of holes is introduced into YIG,  
where iron (Fe) is filled inside every hole. 
Chiral spin-wave edge modes are propagating 
along the boundary in a unidirectional way 
(light purple arrow).}
\end{center}
\label{fig:MC}
\end{figure}

In the next section, we introduce a topological Chern 
number in these volume-mode bands with a 
band gap. In sec.~III, we consider 
a two-dimensional magnonic crystal (MC), where 
the Chern number for the lowest spin-wave volume mode 
takes non-zero integer values in the 
dipolar region. In sec.~IV, we argue that 
the nonzero Chern integer for the lowest volume-mode 
band results in the same integer numbers of topological 
chiral edge modes (surface modes; Fig.~\ref{fig:MC}), 
whose dispersions run across the band gap 
between the lowest volume-mode band and 
the second lowest volume-mode band. The 
existence of the topological chiral edge mode in 
the MC is further justified by a micromagnetic simulation 
in sec.~V.  
The relevant length scale for the magnonic crystal 
turns out to be from sub-$\mu$m to sub-mm, well 
within the range of nanosize fabrication. Unidirectional 
propagations of spin waves along the edge modes 
are experimentally measurable 
especially in yttrium iron garnet (YIG), where the coherence 
length of magnons is on the order of centimeters.~\cite{Serga} 
Based on these observations, we argue in sec.~VI that the 
topological chiral edge modes can be easily 
channelized, twisted, split and 
manipulated, which enables to construct novel magnonic 
devices such as spin-wave logic gate and spin-wave 
current splitter.   

\

\section{Chern number in boson systems} 
\subsection{bosonic BdG Hamiltonian}
To introduce topological Chern number for 
spin-wave volume-mode band, we first consider 
a quadratic form of generic boson Hamiltonian;
\begin{eqnarray}
\hat{\cal H} = \frac12 \sum_{{\bm k}} 
\left[\begin{array}{cc}
{\bm \beta}^{\dag}_{\bm k} & 
{\bm \beta}_{-{\bm k}}\\
\end{array}\right] \cdot {\bm H}_{\bm k} 
\cdot \left[\begin{array}{c}
{\bm \beta}_{\bm k} \\  
{\bm \beta}^{\dag}_{-{\bm k}} \\
\end{array}\right], \label{bH}
\end{eqnarray} 
where ${\bm \beta}^{\dag}_{{\bm k}}\equiv 
[\beta^{\dag}_{1,{\bm k}},\cdots,\beta^{\dag}_{N,{\bm k}}]$ 
denote spin-wave (boson) creation operators. 
Describing volume-type modes, the operators are 
already Fourier-transformed in a two-dimensional 
space with the periodic boundary conditions and 
the wavevector ${\bm k}\equiv (k_x,k_y)$. 
$N$ is a number of internal degrees of 
freedom considered within a unit cell.   
A $2N$ by $2N$ Hermitian 
matrix (${\bm H}_{\bm k}$) stands for a bosonic 
Bogoliubov-de Gennes (BdG) Hamiltonian, whose 
explicit form will be derived from a linearized 
Landau-Lifshitz equation later.  
With the magnetic dipolar interaction,  
the Hamiltonian 
thus derived acquires not only $N$ by $N$ 
normal parts (particle-hole channel), 
${\bm a}_{\bm k}$ and ${\bm a}^{*}_{-{\bm k}}$, but also  
$N$ by $N$ anomalous parts (particle-particle channel), 
${\bm b}_{\bm k}$ and ${\bm b}^{*}_{-{\bm k}}$;   
\begin{eqnarray}
{\bm H}_{\bm k} \equiv 
\left[\begin{array}{cc}
{\bm a}_{\bm k} & {\bm b}_{\bm k} \\
{\bm b}^{*}_{-{\bm k}} & {\bm a}^{*}_{-{\bm k}} \\
\end{array}\right]. \nn
\end{eqnarray}
 
Such a bosonic BdG Hamiltonian is diagonalized in 
terms of a para-unitary matrix ${\bm T}_{\bm k}$ 
instead of a unitary matrix,~\cite{Colpa}  
\begin{eqnarray}  
{\bm T}^{\dag}_{\bm k} \!\ {\bm H}_{\bm k} 
\!\ {\bm T}_{\bm k} = \left[\begin{array}{cc}
{\bm E}_{{\bm k}} & \\
& {\bm E}_{-{\bm k}} \\
\end{array}\right], \label{para}  
\end{eqnarray}  
with $[{\bm \gamma}^{\dag}_{\bm k}, 
{\bm \gamma}_{-{\bm k}}] \!\ {\bm T}^{\dag}_{\bm k} =
[{\bm \beta}^{\dag}_{\bm k}, {\bm \beta}_{-{\bm k}}]$.   
${\bm E}_{{\bm k}}$ is a diagonal matrix,  
whose diagonal element gives a  
dispersion relation for respective volume-mode band. 
The orthogonality and completeness of a 
new basis (${\bm \gamma}$-field) are derived as  
\begin{eqnarray}
{\bm T}_{\bm k}^{\dag} {\bm \sigma}_3 
{\bm T}_{\bm k}
= {\bm \sigma}_3,  \!\ \!\ 
{\bm T}_{\bm k} {\bm \sigma}_3  
{\bm T}^{\dag}_{\bm k} 
= {\bm \sigma}_3 \label{orth-comp}
\end{eqnarray}
respectively, where a diagonal matrix   
${\bm \sigma}_3$ takes $\pm 1$ in the particle/hole 
space, i.e. $[{\bm \sigma}_3]_{jm} = \delta_{jm}\sigma_j$ 
with $\sigma_j=+1$ for $j=1,\cdots,N$ 
and $\sigma_j=-1$ for $j=N+1,\cdots,2N$.  
This additional structure comes from 
the fact that the magnon obeys the boson 
statistics. Each column vector encoded 
in the paraunitary matrix 
${\bm T}_{\bm k}$ stands for (periodic part of)
Bloch wavefunction for the 
respective volume-mode band.  

Provided that a Hermite matrix ${\bm H}_{\bm k}$ 
is unitarily equivalent to a positive-definite diagonal matrix, 
a para-unitary matrix ${\bm T}_{\bm k}$ which diagonalizes 
${\bm H}_{\bm k}$ can be obtained by a 
method based on the Cholesky decomposition.~\cite{Colpa}  
In the method, we first decompose ${\bm H}_{\bm k}$ 
into a product between an upper triangle 
matrix ${\bm K}_{\bm k}$ and its Hermite conjugate; 
${\bm H}_{\bm k} = 
{\bm K}^{\dag}_{\bm k} {\bm K}_{\bm k}$. 
The unitarily positive definiteness of ${\bm H}_{\bm k}$ 
always allows this decomposition and also 
guarantees the existence of ${\bm K}^{-1}_{\bm k}$.
We next introduce a unitary matrix ${\bm U}_{\bm k}$ which 
diagonalizes a Hermite 
matrix ${\bm W}_{\bm k} \equiv 
{\bm K}_{\bm k}{\bm \sigma}_3 {\bm K}^{\dag}_{\bm k}$;
\begin{eqnarray}
{\bm U}^{\dag}_{\bm k} {\bm W}_{\bm k} 
{\bm U}_{\bm k} = \left[\begin{array}{cc}
{\bm E}_{\bm k} & \\
& -{\bm E}_{-{\bm k}} \\
\end{array}\right]. \nn 
\end{eqnarray}
Owing to the Sylvester's law of inertia, 
both ${\bm E}_{\bm k}$ and 
${\bm E}_{-{\bm k}}$ can be made positive-definite 
$N$ by $N$ diagonal matrices. One can see a posteriori 
that these two diagonal matrices are nothing but  
those in the right hand side of Eq.~(\ref{para}). 
Namely, the following paraunitary matrix  
satisfies Eq.~(\ref{orth-comp});~\cite{Colpa} 
\begin{eqnarray}
{\bm T}_{\bm k} = {\bm K}^{-1}_{\bm k} \!\ 
{\bm U}_{\bm k} \!\ \left[\begin{array}{cc} 
{\bm E}^{\frac{1}{2}}_{\bm k} & \\
& {\bm E}^{\frac{1}{2}}_{-{\bm k}} \\
\end{array}\right]. \label{Colpa}
\end{eqnarray}
and it diagonalizes the Hamiltonian as;  
\begin{eqnarray}
{\bm H}_{\bm k} {\bm T}_{\bm k} 
= {\bm \sigma}_3 {\bm T}_{\bm k}\left[\begin{array}{cc}
{\bm E}_{\bm k} & \\
& - {\bm E}_{-{\bm k}} \\
\end{array}\right]. \label{para2}
\end{eqnarray}
The upper $N$ by $N$ diagonal matrix 
in the right hand side, ${\bm E}_{\bm k}$, is positive 
definite, so that we will 
referred to them as (dispersions for) `particle bands', while 
the lower $N$ by $N$ diagonal matrix,  
$-{\bm E}_{-{\bm k}}$, is negative definite, 
whose diagonal elements are thus referred to 
as (dispersion for) `hole bands'. Due to the 
trivial redundancy, ${\bm \sigma}_1 {\bm H}^{*}_{\bm k} 
{\bm \sigma}_1={\bm H}_{-{\bm k}}$ with 
$[{\bm \sigma}_1]_{jm}=\delta_{|j-m|,N}$, either 
one of these two $N$ by $N$ diagonal matrices 
gives the full information of the dispersions for 
the volume-mode bands.

\subsection{Chern integers in bosonic BdG systems}
To introduce the Chern number for 
the $j$-th volume-mode band, let us first 
define a projection operator $P_j$ in the $2N$ dimensional 
vector space, which filters out those bands 
other than the $j$-th volume-mode band at 
each momentum point ${\bm k}$;
\begin{eqnarray}
{\bm P}_{j} \equiv {\bm T}_{\bm k} 
{\bm \Gamma}_j {\bm \sigma}_3 
{\bm T}^{\dag}_{\bm k} {\bm \sigma}_3.  \label{pro}
\end{eqnarray}
Here ${\bm \Gamma}_j$ is a diagonal matrix 
taking $+1$ for the $j$-th diagonal component 
and zero otherwise. Eq.~(\ref{orth-comp}) suggests that 
the operator obeys $\sum_{j} {\bm P}_j = {\bm 1}$ 
and ${\bm P}_j {\bm P}_m= \delta_{jm} {\bm P}_{j}$. 
In terms of the projection operator, the Chern number 
for the $j$-th band is given as 
follows,~\cite{Avron}  
\begin{align}
C_j \equiv \frac{i\epsilon_{\mu\nu}}{2\pi} \int_{\rm BZ} 
d{\bm k} \!\ 
{\rm Tr}\big[({\bm 1}-{\bm P}_j) 
\big(\partial_{k_{\mu}}{\bm P}_j\big)
\big(\partial_{k_{\nu}}{\bm P}_j\big)\big],   
 \label{Ch1}
\end{align}  
where the integral is over the first Brillouin 
zone (BZ) in the two-dimensional 
${\bm k}$ space. 

Eq.~(\ref{Ch1}) is  
integer-valued and characterizes a 
certain global phase structure associated with 
a Bloch wavefunction over the 
BZ. To see this, we follow  
the same argument as in the quantum 
Hall case,~\cite{TKNN,Ko} and  
introduce field strength (Berry's curvature) 
$B_j$ and  gauge connection (gauge field) 
$(A_{j,x},A_{j,y})$ for each volume-mode band; 
\begin{align}
&B_{j}({\bm k}) \equiv  
\partial_{k_x} {A}_{j,y}({\bm k}) 
- \partial_{k_y} {A}_{j,x}({\bm k}),
\label{Ch3} \\
&{A}_{j,\nu}({\bm k}) \equiv  
i{\rm Tr}[{\bm \Gamma}_j {\bm \sigma}_3  
{\bm T}^{\dag}_{\bm k}{\bm \sigma}_3
(\partial_{k_{\nu}}{\bm T}_{\bm k})], \label{Ch4} 
\end{align}
with $j=1,\cdots,2N$. The Chern number for a volume-mode 
band reduces to an integral of the respective 
Berry's curvature over the BZ,      
\begin{align}
C_j = \frac{1}{2\pi} \!\ 
\int_{{\rm BZ}} d^2{\bm k} \!\ 
B_{j}({\bm k}). \label{Ch2} 
\end{align}
Such a surface integral is zero, 
provided that the respective gauge field can be 
defined uniquely and smoothly over the first BZ. 
When $[{\bm T}_{\bm k}]_{m,j}$ 
has a zero somewhere on the BZ for 
any $m$ ($m=1,\cdots,2N$), however, 
the gauge field for the $j$-th band cannot be chosen  
uniquely over the BZ. In this case, 
it is necessary to decompose the BZ  
into two overlapped regions ($H_{1}$ and $H_{2}$ with 
$H_{1} \cup H_{2} = {\rm BZ}$ and $H_{1} \cap H_{2} = 
\partial H_{1} = - \partial H_{2} \equiv S$), so 
that $[{\bm T}_{\bm k}]_{1,j}$ does not have 
any zero within one region ($H_{1}$), while 
$[{\bm T}_{\bm k}]_{2,j}$ has no  
zero inside the other ($H_{2}$). In the former 
region, we then take the gauge, 
$[{\bm T}^{(1)}_{\bm k}]_{m,j}$, such that  
$[{\bm T}^{(1)}_{\bm k}]_{1,j}$ is always real 
positive, while take another  
gauge in the 
other, making 
$[{\bm T}^{(2)}_{\bm k}]_{2,j}$   
to be always real positive. Provided that 
the $j$-th band considered 
is isolated from the others 
($E_{{\bm k},j} \ne 
E_{{\bm k},m \ne j}$ for any ${\bm k}$), 
these two gauge choices are 
related to each other by a  
$U(1)$ transformation,   
\begin{align}
{\bm T}^{(2)}_{\bm k} {\bm \Gamma}_j =&  
 {\bm T}^{(1)}_{\bm k} {\bm \Gamma}_j 
\!\ e^{i\theta_{\bm k}}, 
\label{freedom} 
\end{align} 
on ${\bm k} \in H_1 \cap H_2$. 
Now that the gauge of 
${\bm T}^{(1)}_{\bm k}{\bm \Gamma}_j$
and the gauge of ${\bm T}^{(2)}_{\bm k}{\bm \Gamma}_j$ 
are uniquely defined in $H_{1}$ and 
$H_{2}$ respectively, 
\begin{align}
{A}^{(m)}_{j,\nu} \equiv i{\rm Tr}[{\bm \Gamma}_j {\bm \sigma}_3  
{{\bm T}^{(m)}_{\bm k}}^{\dag}{\bm \sigma}_3
(\partial_{k_{\nu}} {\bm T}^{(m)}_{\bm k})] \label{gauge-dif}
\end{align}
($m=1,2$) are smooth functions in each of these two regions 
respectively. 
The Stokes theorem is applied separately, so that  
Eq.~(\ref{Ch2}) is calculated as, 
\begin{align}
C_j = \frac{1}{2\pi} \!\ \oint_{S} d{\bm k} 
\cdot \big({A}^{(1)}_{j} - {A}^{(2)}_j \big) 
= \frac{1}{2\pi} \oint_{S} d{\bm k} 
\cdot {\nabla}_{\bm k} \theta_{\bm k}, \label{Ch4a}  
\end{align} 
with ${\nabla}_{\bm k} \equiv (\partial_{k_x},\partial_{k_y})$. 
Two regions share a boundary $(S)$, 
which forms a closed loop. $\theta_{\bm k}$ in 
Eq.~(\ref{freedom}) has a $2\pi n$ phase 
winding along the loop. This leads 
to $C_j=n$ ($n = {\mathbb Z}$).

\subsection{Sum rule for Chern integer} 
When all volume-mode bands in a  
system are physically stable, the sum 
of the Chern integer over all particle 
bands and that over all hole bands
are zero respectively;
\begin{eqnarray}
\sum^{N}_{j=1} C_j =\sum^{2N}_{j=N+1} C_j = 0. \label{sum-rule}
\end{eqnarray}   
To see this, let us linearly interpolate a $2N$ by $2N$  
spin-wave Hamiltonian and the $2N$ by $2N$ unit 
matrix; 
\begin{eqnarray}
{\bm H}_{{\bm k},\lambda}  
=(1-\lambda) {\bm H}_{\bm k} + 
\lambda {\bm 1} \label{linearly}
\end{eqnarray} 
We assume that ${\bm H}_{\bm k}$ is paraunitarily equivalent 
to a diagonal matrix whose elements are all positive 
for any wavevector ${\bm k}$; all volume-mode 
bands obtained from original spin-wave Hamiltonian 
($\lambda=0$) are physically stable.  
Thanks to the Sylvester's law of inertia,  
such an Hermite matrix is unitarily equivalent 
to a diagonal matrix whose elements are all positive 
definite. Clearly, so is any ${\bm H}_{{\bm k},\lambda}$ 
during $0\le \lambda \le 1$. Given the unitarily positive 
definiteness, we can then introduce from Eq.~(\ref{Colpa})
a paraunitary matrix ${\bm T}_{{\bm k},\lambda}$ 
which transforms ${\bm H}_{{\bm k},\lambda}$ into 
a diagonal form as;
\begin{align}
{\bm H}_{{\bm k},\lambda} 
{\bm T}_{{\bm k},\lambda} = {\bm \sigma}_3 
{\bm T}_{{\bm k},\lambda}
\left[\begin{array}{cc}
{\bm E}_{{\bm k},\lambda} & \\
& - {\bm E}_{-{\bm k},\lambda} \\
\end{array}\right], \nn
\end{align}  
with positive-definite $N$ by $N$ diagonal matrices  
${\bm E}_{{\bm k},\lambda}$ and ${\bm E}_{-{\bm k},\lambda}$. 
With ${\bm T}_{{\bm k},\lambda}$, 
the Chern integer can be explicitly defined 
as a function of $\lambda$ for $0\le \lambda \le 1$,  
$C_j(\lambda)$. 

The sum of Chern integer 
over a group of bands does not change, 
unless some band in the group forms a band 
touching (frequency degeneracy) with bands outside 
the group. The positive definiteness of 
${\bm E}_{{\bm k},\lambda}$ and ${\bm E}_{-{\bm k},\lambda}$ 
means that particle bands obtained 
from ${\bm H}_{{\bm k},\lambda}$ are always in positive 
frequency regime, while hole bands are in negative frequency 
regime; during the interpolation ($0\le \lambda \le 1$), they 
are always disconnected from each other in frequency.  
Thus, the sum of the Chern integer 
over all particle bands does not change 
during the interpolation,
\begin{align}
\sum^{N}_{j=1} C_j (\lambda=0) = 
\sum^{N}_{j=1} C_j (\lambda=1). \nn 
\end{align} 
Since a paraunitary matrix at $\lambda=1$ is trivial, 
${\bm T}_{{\bm k},\lambda=1} = 1$, the right hand side 
reduces to zero and so is the case with the original 
spin-wave Hamiltonian ($\lambda=0$). In summary, 
Eq.~(\ref{sum-rule}) is derived 
only from the para-unitarily positive definiteness 
of $2N$ by $2N$ Hermite matrix ${\bm H}_{\bm k}$. 
 As a corollary of eq.~(\ref{sum-rule}), one 
can argue that any topological chiral edge modes 
obtained from proper spin-wave 
approximations appear only at a {\it finite} frequency 
region (see the sec.~IV for the argument).

In the following, we show that 
a two-dimensional bicomponent magnonic crystal (MC) 
with the dipolar interaction supports 
spin-wave bands with non-zero Chern integers.

\section{magnonic crystal, dipolar interaction 
and Chern integer for volume-mode bands}
\subsection{plane-wave theory for magnonic crystals}
The MC considered is a ferromagnetic system with 
its magnetization and exchange interaction  
modulated periodically in the 2-dimensional 
($x$-$y$) direction. For simplicity, we assume that 
the system is translationally symmetric along the 
$z$-direction, 
whereas the subsequent results are
expected to be similar when a thickness of the 
system in the $z$-direction becomes finite (but 
large). 
The system is composed 
of two kinds of ferromagnets; iron and YIG. 
The unit cell of the MC is an $a_x \times a_y$ 
rectangle, inside which iron is embedded 
into circular regions, while the 
remaining region is filled with YIG (Fig.~\ref{fig:MC}).  
A uniform magnetic field $H_0$ is applied 
along the $z$ direction, such that
the static ferromagnetic moment $M_s$ in 
both regions is fully polarized in the longitudinal 
direction. Propagation of the transverse moments $(m_x,m_y)$ 
is described by a linearized 
Landau-Lifshitz 
equation~\cite{Kalinikos,MC1,MC2,MC3}; 
\begin{align}
\frac{1}{|\gamma|\mu_{0}} 
\frac{dm_{\pm}}{dt} = &\pm 2i M_s 
\big(\nabla\cdot Q \nabla\big) m_{\pm} 
\mp 2im_{\pm} \big(\nabla\cdot Q \nabla\big) M_s 
\nn \\ 
&\hspace{1cm}
\mp i H_0 m_{\pm} \pm i h_{\pm} M_s  \label{LL}
\end{align}
with $\nabla\equiv (\partial_x,\partial_y)$, 
$m_{\pm}= m_{x}\pm im_y$ and 
$h_{\pm}=h_x\pm ih_y$. $(h_x,h_y)$ 
stands for transverse component of 
long-ranged magnetic dipolar field ${\bm h}$, 
which is related to the ferromagnetic moment 
${\bm m} \equiv (m_x,m_y,M_s)$ via the 
Maxwell equation, i.e. 
$\nabla \times {\bm h}= c^{-1}\partial_{t} e_z$ 
and $\nabla \cdot ({\bm h} + {\bm m})=0$. 
The former two terms in the right hand side 
of Eq. (\ref{LL}) come from short-ranged 
exchange interaction, where $Q$ denotes a  
square of the exchange interaction length.  
$M_s$ and $Q$ take values of iron inside  
the circular region and those of  
YIG otherwise. A filling fraction of the 
circular region with respect to the total area 
of the unit cell is represented by $f$. 
We further employ magneto-static 
approximation, replacing the Maxwell equations by 
$\nabla \times {\bm h}=0$ and 
$\nabla \cdot ({\bm h} + {\bm m})=0$;  
\begin{eqnarray}
h_{\nu} = - \partial_{\nu} \Psi, \ \ 
\Delta \Psi = \partial_x m_x + \partial_y m_y,  
\label{mstatic}  
\end{eqnarray}
with $\nu=x,y$. 
This in combination with Eq.~(\ref{LL}) gives 
a closed equation of motion (EOM) for the 
transverse moment. 

In order to obtain band dispersions 
and Chern integers for volume-mode bands, 
we need to reduce the EOM into a generalized 
eigenvalue problem with a BdG Hamiltonian 
(${\bm H}_{\bm k}$) defined in the form of Eq.~(\ref{bH}). 
To this end, we first normalize the transverse moment, 
to introduce  a Holstein-Primakov (HP) field as;~\cite{Mag} 
\begin{eqnarray}
\beta({\bm r}) \equiv  
\frac{m_{+}({\bm r})}{\sqrt{2M_s({\bm r})}}, \!\ 
\!\  
\beta^{\dag}({\bm r}) \equiv  
\frac{m_{-}({\bm r})}{\sqrt{2M_s({\bm r})}}.  \label{HP1}
\end{eqnarray} 
In terms of the HP field,  
the Landau-Lifshitz equation is 
properly symmetrized as,   
\begin{widetext}
\begin{align}
\frac{d\beta}{dt} = & 4i \alpha \big(\nabla \cdot Q \nabla\big) 
\alpha \beta - 4i \beta\big(\nabla \cdot Q\nabla\big) \alpha^2
 - iH_0 \beta - i \alpha \partial_{+} \Psi, \label{eom1} \\
\Delta \Psi = & \partial_+ \big(\alpha\beta^{\dag}\big) + \partial_{-} 
\big(\alpha \beta\big), \ \ \ \partial_{\pm}\equiv \partial_x\pm i\partial_y,  
\label{eom2}
\end{align}  
\end{widetext}
with $\alpha({\bm r})\equiv \sqrt{M_s ({\bm r})/2}$. 
$|\gamma|\mu_0$ was omitted in Eq.~(\ref{eom1}) 
for clarity. The static magnetization and exchange interaction 
are spatially modulated 
with the lattice periodicity; 
\begin{align}
\alpha({\bm r}) = \sum_{{\bm G}}
\alpha({\bm G}) \!\ e^{i{\bm G}\cdot {\bm r}}, \ \ 
Q({\bm r})= \sum_{{\bm G}}
Q({\bm G}) \!\ e^{i{\bm G}\cdot {\bm r}}, \nn 
\end{align}
with the reciprocal vectors ${\bm G}$, 
$\alpha^{*}({\bm G})=\alpha(-{\bm G})$ 
and $Q^{*}({\bm G})=Q(-{\bm G})$. It 
follows from the Bloch theorem that 
${\bm m}({\bm r})$ and $\Psi({\bm r})$ take 
a form;   
\begin{align}
{\beta}({\bm r}) 
& = \sum_{\bm k}\sum_{{\bm G}} {\beta}_{\bm k}({\bm G}) 
\!\ e^{i({\bm k}+{\bm G})\cdot {\bm r}}, \nn \\
{\beta}^{\dagger}({\bm r}) 
& = \sum_{\bm k}\sum_{{\bm G}} {\beta}^{\dagger}_{\bm k}({\bm G}) 
\!\ e^{-i({\bm k}+{\bm G})\cdot {\bm r}}, \nn 
\end{align}
and 
\begin{align}
\Psi({\bm r}) & = 
\sum_{\bm k}\sum_{{\bm G}} {\Psi}_{\bm k}({\bm G}) 
\!\ e^{i({\bm k}+{\bm G})\cdot {\bm r}}, \nn
\end{align} 
where the ${\bm k}$-summation  
is taken over the first BZ. 

In terms of these Fourier modes, an equivalent 
generalized eigenvalue problem with a 
quadratic Hamiltonian for the HP field is  
derived as
\begin{align}
i\frac{d}{dt} \left[\begin{array}{c}
{\bm \beta}_{\bm k} \\
{\bm \beta}^{\dag}_{-{\bm k}} \\
\end{array}\right] 
&= \big[\left[\begin{array}{c} 
{\bm \beta}_{\bm k} \\
{\bm \beta}^{\dag}_{-{\bm k}} \\
\end{array}\right], \hat{\cal H} \!\ \big] 
= {\bm \sigma}_3 {\bm H}_{\bm k} 
\left[\begin{array}{c}
{\bm \beta}_{\bm k} \\
{\bm \beta}^{\dagger}_{-{\bm k}} \\
\end{array}\right], \label{eom3a} 
\end{align} 
with 
\begin{align}
\hat{\cal H} &= \sum_{k_y>0} 
\left[\begin{array}{cc}
{\bm \beta}^{\dag}_{\bm k} & 
{\bm \beta}_{-{\bm k}}\\
\end{array}\right] {\bm H}_{\bm k} 
\left[\begin{array}{c}
{\bm \beta}_{\bm k} \\  
{\bm \beta}^{\dag}_{-{\bm k}} \\
\end{array}\right], \nn 
\end{align}
and 
\begin{align}
\left[\begin{array}{c}
{\bm \beta}_{\bm k}  \\ 
{\bm \beta}^{\dagger}_{-{\bm k}} \\
\end{array}\right]  
&\equiv [\cdots,\beta_{\bm k}({\bm G}),\cdots, 
\beta_{\bm k}(-{\bm G}),\cdots,| \nn \\
&\hspace{0.5cm} 
\cdots,\beta^{\dagger}_{-{\bm k}}(-{\bm G}),\cdots
\beta^{\dagger}_{-{\bm k}}({\bm G}),\cdots]^T, \nn \\ 
%
\left[\begin{array}{cc}
{\bm \beta}^{\dagger}_{\bm k} & 
{\bm \beta}_{-{\bm k}} \\
\end{array}\right] 
&\equiv [\cdots,\beta^{\dagger}_{\bm k}({\bm G}),\cdots
\beta^{\dagger}_{\bm k}(-{\bm G}),\cdots,| \nn \\
& \hspace{0.5cm} 
\cdots,\beta_{-{\bm k}}(-{\bm G}),\cdots
\beta_{-{\bm k}}({\bm G}),\cdots]. \nn  
\end{align}
${\bm \sigma}_3$ in the right hand side 
takes $\pm 1$ in the particle/hole space,
which comes from the  
commutation relation of bosons, 
$[{\bm \beta}_{\bm k},
{\bm \beta}^{\dag}_{{\bm k}}] = {\bm 1}$ 
and $[{\bm \beta}^{\dag}_{-{\bm k}},
{\bm \beta}_{-{\bm k}}] = -{\bm 1}$.   
A comparison between Eqs~(\ref{eom1},\ref{eom2}) 
and Eq.~(\ref{eom3a}) dictates that 
${\bm H}_{\bm k}$ thus introduced is 
given by a following Hermitian matrix 
(${\bm H}^{\dag}_{\bm k}
= {\bm H}_{\bm k}$);  
\begin{eqnarray}
{\bm H}_{\bm k} 
\equiv \left[\begin{array}{cc}
{\bm \alpha} \cdot {\bm \alpha} + {\bm B}_{\bm k} + H_0 {\bm 1} 
& {\bm \alpha} \cdot {\bm I}_{\bm k} \cdot {\bm \alpha} \\
{\bm \alpha} \cdot {\bm I}^{*}_{\bm k} \cdot {\bm \alpha}  
& {\bm \alpha} \cdot {\bm \alpha} + {\bm B}_{\bm k} + H_0 {\bm 1}  \\
\end{array}\right] \label{bdg}
\end{eqnarray}
with 
\begin{align}
[{\bm \alpha}]_{{\bm G},{\bm G}'} &\equiv 
\alpha({\bm G}-{\bm G}'), \ \  
[{\bm I}_{\bm k}]_{{\bm G},{\bm G}'} 
\equiv \delta_{{\bm G},{\bm G}'} 
e^{-2i\theta_{\bm k}({\bm G})}, \nn \\
e^{i\theta_{\bm k}({\bm G})} &\equiv 
\frac{({\bm k}+{\bm G})_x + 
i ({\bm k}+{\bm G})_y}{|{\bm k}+{\bm G}|}, \label{phase} 
\end{align} 
and 
\begin{widetext}
\begin{align}
[{\bm B}_{\bm k}]_{{\bm G},{\bm G}'} 
&\equiv 4 \sum_{{\bm G}_1,{\bm G}_2} 
\alpha({\bm G}-{\bm G}_1) Q({\bm G}_1-{\bm G}_2)  
\alpha({\bm G}_2-{\bm G}') \!\ 
({\bm k}+{\bm G}_1) \cdot ({\bm k}+{\bm G}_2) \nn \\
& -4  \sum_{{\bm G}_1,{\bm G}_2} 
Q({\bm G}_1) 
\alpha({\bm G}_2) 
\alpha({\bm G}-{\bm G}'-{\bm G}_1-{\bm G}_2) \!\ 
({\bm G}-{\bm G}') \cdot 
({\bm G}-{\bm G}'-{\bm G}_1). \label{Dk}
\end{align}
\end{widetext}
After taking the summation over ${\bm G}_1$ 
and ${\bm G}_2$ in Eq.~(\ref{Dk}), one 
can further decompose 
$[{\bm B}_{\bm k}]_{{\bm G},{\bm G}'}$ 
into three parts,
\begin{align}
[{\bm B}_{\bm k}]_{{\bm G},{\bm G}'}
= & \!\ 4 \sum_{\mu=x,y} 
\big[Q \alpha^2\big]_{{\bm G}-{\bm G}'} \!\ 
({\bm k}+{\bm G})_{\mu} 
({\bm k} +{\bm G}')_{\mu} \nn \\
& \!\ \!\ + 4 i \sum_{\mu=x,y}
\big[Q \alpha (\partial_{\mu} \alpha) \big]_{{\bm G}-{\bm G}'} 
\!\ ({\bm G}-{\bm G}')_{\mu} \nn \\
&\!\ \!\ \!\ \!\  + 4 \sum_{\mu=x,y} 
\big[Q (\partial_{\mu} \alpha) 
(\partial_{\mu} \alpha) \big]_{{\bm G}-{\bm G}'} \label{Dka}
\end{align}
with 
\begin{align}
\big[Q \alpha^2\big]_{\bm G} &\equiv 
\frac{1}{S} \int  
Q ({\bm r}) \alpha^2({\bm r}) \!\ 
e^{-i {\bm G}\cdot {\bm r}}  d^2{\bm r}, \nn  \\
\big[Q \alpha(\partial_{\mu} \alpha)\big]_{\bm G} &\equiv 
\frac{1}{S} \int  
Q({\bm r}) \alpha({\bm r}) 
(\partial_{\mu} \alpha({\bm r})) \!\ 
e^{-i {\bm G}\cdot {\bm r}} \!\ d^2{\bm r}, \nn  \\
\big[Q(\partial_{\mu} \alpha) 
(\partial_{\mu} \alpha)\big]_{\bm G} &\equiv 
\frac{1}{S} \int  
Q({\bm r}) (\partial_{\mu} \alpha({\bm r}))  
(\partial_{\mu} \alpha({\bm r})) \!\ 
e^{-i {\bm G}\cdot {\bm r}} \!\ d^2{\bm r}, 
\label{3exp}
\end{align}
where the 2-$d$ integrals in the right 
hand side are taken over the MC unit cell and 
$S$ denotes an area of the cell ($S \equiv a_xa_y$). 
In actual numerical calculation, 
the dimension of ${\bm H}_{\bm k}$ is 
typically taken to be $512 \times 512$, 
where the reciprocal vector ${\bm G}$ 
ranges over 
$[-16\pi/a_x,16\pi/a_x]\times [-16\pi/a_y,16\pi/a_y]$.

The MC considered is composed of two 
kinds of ferromagnets, where  
the respective (square root of) 
magnetization and (square of) 
exchange interaction length are specified 
by $(\alpha_j,Q_j)$ $(j=1,2)$. 
Within the rectangular-shaped unit cell ($a_x \times a_y$), 
one of these two ferromagnets $(\alpha_1,Q_1)$ 
is embedded within the circular region, while 
the remaining region is filled with the other 
$(\alpha_2,Q_2)$. 
If $\alpha({\bm r})$ has a discontinuity 
at the boundary between these two regions, 
the last term in Eq.~(\ref{Dka}), 
$[Q(\partial_{\mu}\alpha)(\partial_{\mu}\alpha)]$, 
diverges, since it contains   
the second derivative with 
respect to a spatial coordinate along 
the radial direction. 
Physically, such an infrared divergence  
is removed by a smooth variation of 
the saturation magnetization at the boundary. 
For simplicity, we interpolate 
$\alpha({\bm r})$ as a linear function 
of the radial coordinate measured 
from the center of the circular region;
\begin{eqnarray}
\alpha({\bm r})=\left\{\begin{array}{ll}
\alpha_1 & (|{\bm r}|<R_0) \\
\alpha_1 - \frac{|{\bm r}|-R_0}{R_1-R_0} 
(\alpha_1 - \alpha_2)  
& (R_0 < |{\bm r}| < R_1) \\
\alpha_2 &  (R_1<|{\bm r}|) \\
\end{array}. \right.  
\end{eqnarray}   
A discontinuity in $Q({\bm r})$ is 
also removed by the same linear interpolation. 
Actual numerics are carried out with $(r_0,r_1) \equiv 
(R_0/\lambda,R_1/\lambda) = (0.10,0.125)$, 
where $\lambda$ denotes the linear dimension 
of the unit cell size, $\lambda\equiv \sqrt{a_x a_y}$.  
As for the material parameter, we used 
$(M_{s,1},M_{s,2})= (1.8,0.19) \!\ [{\rm A}/\mu{\rm m}]$ 
and $(\sqrt{Q_1},\sqrt{Q_2}) =  
(33,130) \!\ [\AA]$ with $\alpha_j \equiv \sqrt{M_{s,j}}$, while 
parameters of iron (Fe), cobalt (Co), 
and YIG are  
$(M_s\!\ [{\rm A}/\mu{\rm m}],\sqrt{Q} \!\ 
[\AA])=(1.7,21),(1.4,29)$, and $(0.14,184)$ 
respectively. 

Under ${\bm T}_{\bm k}$, 
the EOM is para-unitarily equivalent to
\begin{eqnarray}
i\frac{d}{dt}\left[\begin{array}{c}
{\bm \gamma}_{\bm k} \\
{\bm \gamma}^{\dag}_{-{\bm k}} \\
\end{array}\right] = 
\left[\begin{array}{cc} 
{\bm E}_{{\bm k}} & \\
& - {\bm E}_{-{\bm k}} \\
\end{array}\right] \left[\begin{array}{c}
{\bm \gamma}_{\bm k} \\
{\bm \gamma}^{\dag}_{-{\bm k}} \\
\end{array}\right]. \nn
\end{eqnarray} 
$[{\bm E}_{{\bm k}}]_{j}$ gives a dispersion relation 
for the $j$-th volume-mode band ($j=1,\cdots$), while 
the Chern integer is calculated from ${\bm T}_{\bm k}$ 
via Eqs.~(\ref{Ch3},\ref{Ch4},\ref{Ch2}). 
In its numerical evaluation, we employed an 
algorithm based on a `manifestly gauge-invariant' 
description of the Chern integer.~\cite{FHS}  

\subsection{role of 
dipolar interaction}
When a system considered is either 
time-reversal symmetric; 
${\bm H}^{*}_{-{\bm k}}= {\bm H}_{\bm k}$, 
or mirror-symmetric with 
a mirror plane perpendicular 
to the $xy$ plane, 
e.g. ${\bm H}_{(k_x,k_y)} = {\bm H}_{(k_x,-k_y)}$, 
the Berry's curvature satisfies $-B_j({\bm k})=B_j(-{\bm k})$
or $-B_j(k_x,k_y)= B_j(k_x,-k_y)$, respectively, 
which reduces the Chern integer to zero. 
In the present situation, however, 
the magnetic dipolar field brings about 
complex-valued phase factors in the anomalous 
part, ${\bm I}_{\bm k} \ne {\bm I}^{*}_{\bm k}$, 
which removes from Eq.~(\ref{bdg}) 
both the time-reversal symmetry and 
the mirror symmetries.  
Without periodic modulation of the saturation 
magnetization, ${\bm \alpha} = \alpha{\bm 1}$, 
these phase factors can be erased by a proper gauge 
transformation, ${\bm \beta}^{\dag}_{\bm k}
\rightarrow {\bm \beta}^{\dag}_{\bm k} 
\sqrt{{\bm I}^{*}_{\bm k}}$ and 
${\bm \beta}_{-{\bm k}}
\rightarrow {\bm \beta}_{-{\bm k}} 
\sqrt{{\bm I}_{\bm k}}$, so that both 
symmetries are recovered. In the 
presence of the periodic modulations, 
${\bm \alpha} \ne \alpha{\bm 1}$,  
however, these two symmetries are generally absent 
in the 2-$d$ MCs and the bosonic Chern integer 
can take a non-zero integer value.
  
This situation is quite analogous to what the 
relativistic spin-orbit interaction does in  
ferromagnetic metals.~\cite{Karplus,Onoda} 
Moreover, contrary to the 
spin-orbit interaction, 
a strength of the dipolar interaction  
is an experimentally tunable parameter 
in MCs.~\cite{Lenk} 
When characteristic length scale of  
MC (linear dimension of the 
unit cell size $\lambda \equiv \sqrt{a_x a_y}$) 
becomes larger than the typical exchange length $\sqrt{Q}$,
the dipolar interaction is expected to prevail 
over the exchange interaction.   
 
\begin{figure}
\includegraphics[scale=0.115]{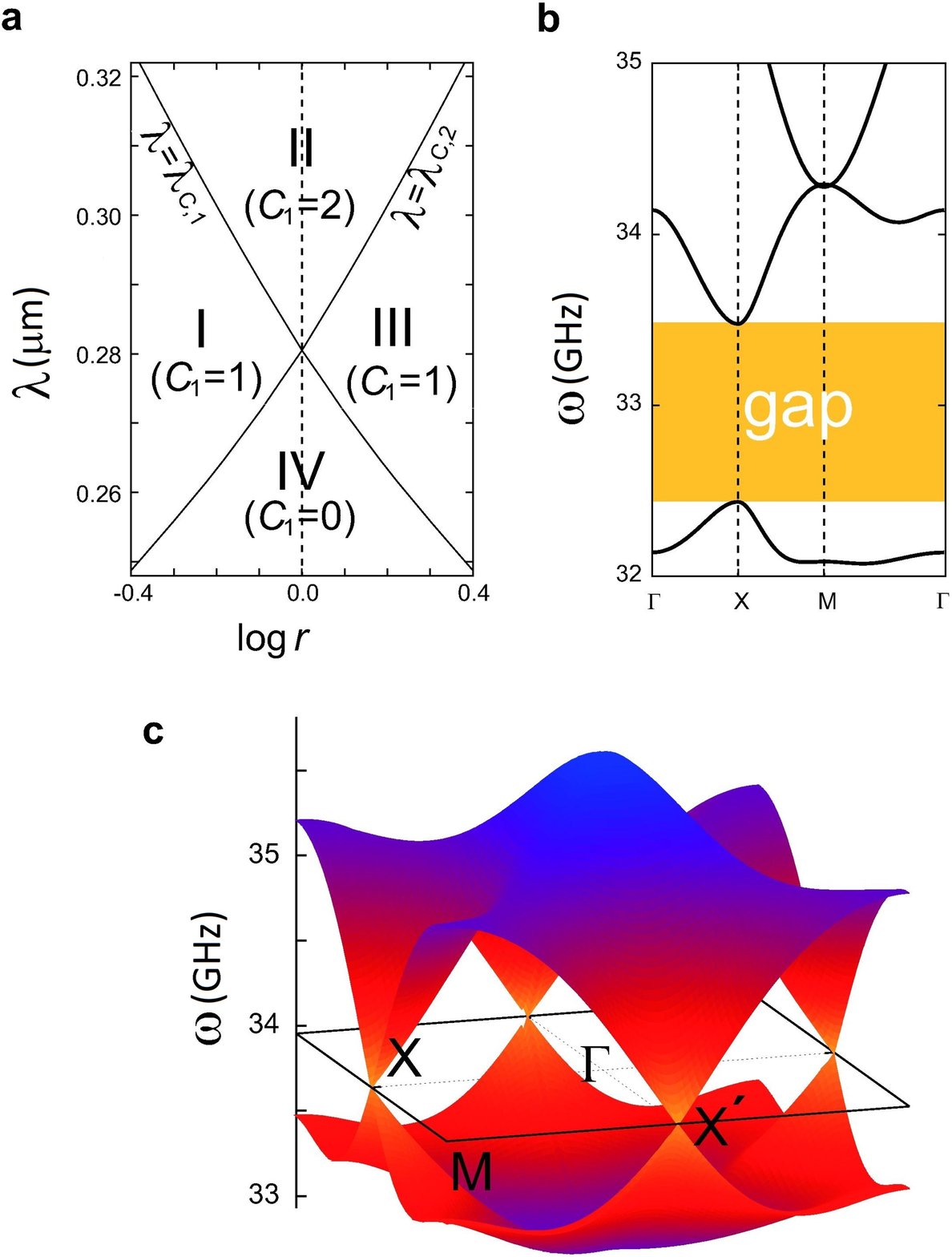}
\caption{Chern-integer phase diagram and 
band dispersions with $f=\pi \times 10^{-2}$. 
\textbf{a,} Chern-integer phase diagram.  
The phases are distinguished by the Chern 
integer of the lowest magnonic band, $C_1$. 
$r$ stands for the aspect ratio of the unit cell 
shape ($r \equiv a_y/a_x$). \textbf{b,} Band dispersions of 
the lowest three volume-mode bands with $r=1$, and 
$\lambda = 0.35 \mu $m.  
A band gap appears between 
the first and the second lowest band. c. 
The band gap  collapses at 
$\lambda=0.28\mu$m ($r=1$), where the 
lowest and second lowest volume mode 
form Dirac cones at the two inequivalent X points.}
\label{fig:MCband}
\end{figure}

\

\
\subsection{Chern integer of volume-mode bands 
and role of band touchings}
\begin{figure}
\includegraphics[scale=0.131]{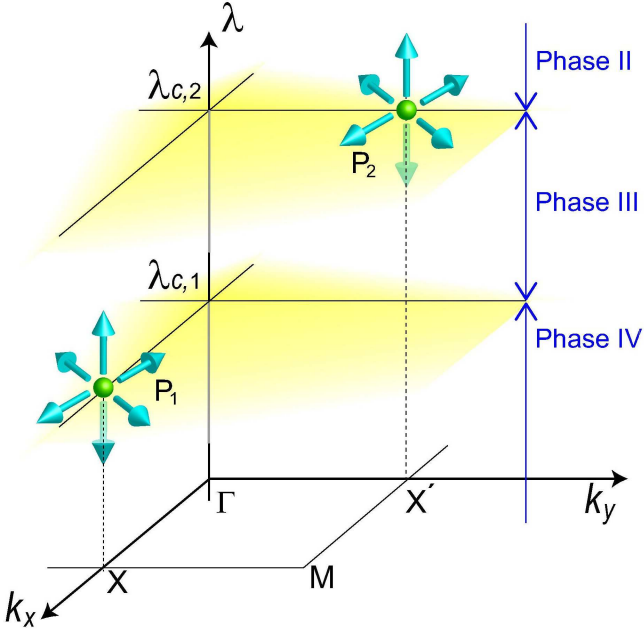}
\caption{Band touching points 
(dual magnetic charges) in a 3-dimensional 
parameter space subtended by the 
wavevector $(k_x,k_y)$ 
and the unit cell size $(\lambda \equiv \sqrt{a_xa_y})$ 
for $r >1$.  Small green spheres denote the 
band touching points, which emit the 
dual magnetic field (blue arrows).} 
\label{fig:monopole}
\end{figure}

In fact, we found that the Chern integer of 
the lowest magnonic band, $C_1$, 
is always quantized to 
be $2$ for the longer $\lambda$, 
while the integer reduces to zero for the 
shorter $\lambda$ (Fig.~\ref{fig:MCband}a).  
The respective quantization is 
protected by a finite direct band gap between 
the lowest band and the second lowest band 
(Fig.~\ref{fig:MCband}b). In the intermediate 
regime of $\lambda$,   
these two bands get closer to each other. 
With a four-fold rotational symmetry 
($r\equiv a_y/a_x = 1$),  the gap closes 
at the two $X$-points at a critical value of 
$\lambda$ ($\sim 0.28\mu$m),  where 
the two bands form gapless 
Dirac spectra 
(Fig.~\ref{fig:MCband}c). 
Without the four-fold symmetry ($r\ne 1$), 
the band touching at one of the two $X$ points  
and that of the other occur at 
different values of $\lambda$. 
These band-touchings 
are denoted as ${\rm P}_1(\pi,0,\lambda_{c,1})$ and 
${\rm P}_2(0,\pi,\lambda_{c,2})$ in a  
3-dimensional parameter space subtended by 
two wavevectors $k_x$, $k_y$ and 
the unit cell size $\lambda$ 
(Fig.~\ref{fig:monopole}).   

The band touchings endow  
the lowest volume-mode band with  
non-zero Chern integers  
in the longer $\lambda$ region. 
In generalized eigenvalue problems 
as well as usual eigenvalue problems,~\cite{TKNN,Ko,Berry,Simon}   
a band-touching point in the $3$-$d$ 
parameter space plays role of a dual magnetic 
monopole (charge). The corresponding 
dual magnetic field 
is generalized from Eq.~(\ref{Ch3})  as 
a rotation of three component gauge field 
${\bm A}_j = (A_{j,x},A_{j,y},A_{j,\lambda})$;  
\begin{eqnarray}
{\bm B}_{j} = {\bm \nabla} \times {\bm A}_j \label{25}
\end{eqnarray}
with ${\bm \nabla} \equiv 
(\partial_{k_x},\partial_{k_y},\partial_{\lambda})$. 
Here the third component of the gauge 
field $A_{j,\lambda}$ is introduced 
in the same way as in (\ref{Ch4});  
\begin{align}
{A}_{j,\lambda} \equiv  
i{\rm Tr}[{\bm \Gamma}_j {\bm \sigma}_3  
{\bm T}^{\dag}_{\bm k}{\bm \sigma}_3
({\partial}_{\lambda}{\bm T}_{\bm k})]. \label{gauge}
\end{align}
$j$ specifies either one of the two 
magnonic bands which form the band-touching. 
At the band-touching point, 
the dual magnetic field for the respective 
bands has a dual magnetic charge, 
whose strength is quantized to be $2\pi$ times 
integer (see Appendix). 
A numerical evaluation tells that the dual 
magnetic charges for the lowest band at the band 
touching points at ${\rm P}_1$ and 
${\rm P}_{2}$ are both $+2\pi$ 
(Fig.~\ref{fig:monopole}); 
\begin{align} 
{\bm \nabla} \cdot {\bm B}_{1} 
= & 2\pi \delta(\lambda-\lambda_{c,1}) \delta(k_x-\pi)
\delta(k_y) \nn \\ 
& \ \ + 2\pi \delta(\lambda-\lambda_{c,2}) \delta(k_x)
\delta(k_y-\pi),  \nn
\end{align}
where 
$\lambda_{c,1}<\lambda_{c,2}$ 
for $a_y>a_x$ ($r>1$) and 
$\lambda_{c,1}>\lambda_{c,2}$ 
for $a_y<a_x$ ($r<1$).  

Because the Chern integer can be regarded 
as the total dual magnetic flux penetrating  
through the constant $\lambda$ plane 
(see Eq.~(\ref{Ch2})), the Gauss theorem 
suggests that, when $\lambda$ goes across either 
$\lambda=\lambda_{c,1}$ plane 
or $\lambda=\lambda_{c,2}$ plane, the 
Chern integer for the lowest magnonic band 
always changes by unit, e.g.  
\begin{align}
& {C_{1}}|_{\lambda>\lambda_{c,1}} - 
{C_{1}}|_{\lambda_{c,1}>\lambda} 
= 1.  \nn
\end{align} 
Without the four-fold rotational symmetry ($r \ne 1$), 
the two critical values of $\lambda$ 
bound three phases for the lowest 
magnonic band, the phase with $C_1=0$ (phase IV), 
that with $C_1=1$ (phase III or I) and that 
with $C_1=2$ (phase II). In the presence of 
the four-fold rotational symmetry ($r=1$), two band 
touchings occur at the same critical value of 
$\lambda$, where $C_1$ increases by two on increasing 
$\lambda$. 
This leads to 
a phase diagram shown in Fig.~\ref{fig:MCband}a, which 
describes the Chern integer of the lowest magnonic 
band as a function of the unit cell size 
$\lambda$ and the aspect ratio $r$. 

A dual magnetic charge is a quantized topological
object, so that, upon any small change of parameters, 
it cannot disappear by itself. Instead, it only 
moves around in the 3-$d$ parameter space. 
As a result, the global 
structure of the phase diagram depicted in 
Fig.~\ref{fig:MCband}a widely holds true for 
other combinations of material 
parameters.  
For $r=1$ and $f=\pi \times 10^{-2}$, 
we found $\lambda_{c}=0.370 \mu $m 
for iron (circular region) and YIG (host), 
and $\lambda_{c}= 0.372 \mu $m for 
cobalt (circular region) 
and YIG (host). When varying the 
filling fraction for iron (circular region) 
and YIG (host) with $r=1$,  
we found $\lambda_{c}=0.274 \mu $m 
for $f=4 \pi \times 10^{-2} $, and 
$\lambda_{c}= 0.348 \mu $m for 
$f=9 \pi \times 10^{-2} $. 

    
\section{Chiral spin-wave edge mode in MC} 
The chiral phases with non-zero 
Chern integers have chiral spin-wave edge modes, 
which are localized at a boundary with the phase  
with zero Chern integer (phase IV) or the vacuum. 
The edge modes have chiral dispersions which go 
across the band gap between the lowest and 
the second lowest band. 

\begin{figure}
\includegraphics[scale=0.1]{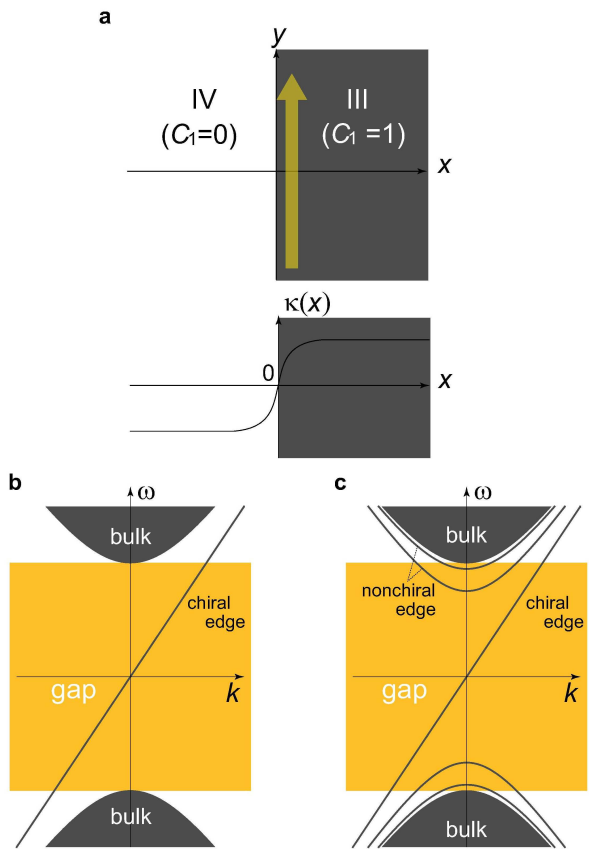}
\caption{Chiral spin-wave edge modes. \textbf{a,} 
geometry of the system. \textbf{bc,} 
wavevector-frequency dispersions for 
Dirac Hamiltonian with the 
P$\ddot{\rm o}$schl-Teller potential~\cite{Hal2} 
$\kappa(x)=\kappa_{\infty}\tanh(x/d)$ 
for $\kappa_{\infty}d=0.9a$ (\textbf{b}) 
and $\kappa_{\infty}d=2.9a$ (\textbf{c}).
The volume-mode magnonic bands have a 
band gap, inside which an edge mode has 
a chiral dispersion.
In \textbf{c} there are some nonchiral 
edge states, whereas in \textbf{b} there 
is not. Nonetheless, 
the number of chiral edge modes is one, 
which is determined solely from the 
difference between the Chern integers 
for the two phases.}
\label{fig:edgemode}
\end{figure}

As an illustrative example, we consider a boundary 
($y$ axis) between the MC in phase III ($C_1=1$) 
and MC in the phase IV ($C_1=0$). The existence of 
a chiral spin-wave edge mode at the boundary is 
shown from a following $2\times 2$ Dirac 
Hamiltonian derived near their phase boundary 
$\lambda=\lambda_{c,1}$ (see appendix), 
\begin{eqnarray}
{\cal H}_{\rm eff} = \omega_0 {\bm \tau}_0 
+  \kappa(x) {\bm \tau}_3 
- i a \partial_{x} {\bm \tau}_1 - 
i b \partial_y {\bm \tau}_2.  \label{effective1}
\end{eqnarray} 
${\bm \tau}_{j}$  denotes 
the Pauli matrices subtended by 
the two-fold degenerate eigenstates at $P_1$. 
$a$ and $b$ are positive material parameters. 
The difference of the Chern integers 
($C_1$) for the two phases
is represented as a change of sign of a 
Dirac mass term  $\kappa(x)$: $\kappa(x)>0$ for $x>0$ 
(phase III) and and $\kappa(x)<0$ 
for $x<0$ (phase IV) (see Fig.~\ref{fig:edgemode}a); 
$\lim_{x\rightarrow \pm \infty}\kappa(x) =\pm 
\kappa_{\infty}$.  The Hamiltonian has a 
following eigenstate;~\cite{Su,Niemi}   
${\bm \psi}_{k}({\bm r}) \propto e^{i k y} 
e^{-\frac{1}{a}\int^{x} 
\kappa(x') dx'} \!\ [1,i]^t$, which 
is localized at the boundary ($x=0$).  
In terms of the surface wavevector $k$ 
along the $y$ axis, the corresponding 
eigen-frequency is given by $E=\omega_0 + bk$. 
This connects the lowest magnonic band lying at 
$E \le \omega_0-\kappa_{\infty}$ and the second lowest band 
at $E \ge \omega_0+\kappa_{\infty}$ 
(see Fig.~\ref{fig:edgemode}b,c).  
The mode is chiral, since the group 
velocity is always positive, 
$v_{k}\equiv\partial_k E=b$. 
Similarly, we can easily show that  
the phase with $C_1=1$ at $r>1$ (Phase III) 
or $r<1$ (Phase I) has a chiral edge mode at its 
boundary with vacuum, whose dispersion crosses 
the direct band gap at the $(\pi,0)$ or $(0,\pi)$-point 
respectively, while 
the phase with $C_1=2$ (Phase II) has both at its  
boundary with vacuum.  A number of 
chiral modes localized at the interface between 
two MCs with different Chern integers 
is equal to the difference of the two Chern integers.

Generalizing the arguments so far 
into the linearly  interpolated Hamiltonian 
defined in eq.~(\ref{linearly}),  
we can argue that, in general, 
a number of those chiral spin-wave edge modes 
whose dispersions run across 
a gap between the $m$-th and the $(m+1)$-th 
particle bands, $N_m$, is equal to the sum 
of the Chern integer over those particle 
bands below the gap;
\begin{align}
N_m \equiv \sum^{m}_{j=1} C_j. \label{general}
\end{align}
Here clockwise (counterclockwise) chiral edge 
modes contribute by $+1$ ($-1$) to the number 
of chiral edge modes, $N_m$ (see Fig.~\ref{fig:edgemode}a). 
Namely, the sum rule, eq.~(\ref{sum-rule}), suggests that 
the right hand side of eq.~(\ref{general}) counts the 
total number of the band touchings (including the sign of the 
respective dual magnetic charges) which happen between 
the $m$-th particle band and $(m+1)$-th particle 
band during the interpolation, 
$\lambda=1 \rightarrow \lambda=0$.   
On the one end, each band touching (dual magnetic monopole) 
is accompanied by the emergence of a chiral edge 
mode between these two bands, whose sense of 
rotation is either clockwise or counterclockwise, depending 
on the sign of the respective dual magnetic charge.   
Since there are no chiral edge modes 
in the trivial limit ($\lambda=1$), we can safely 
conclude eq.~(\ref{general}) at $\lambda=0$. 
As a corollary of eq.~(\ref{general}), one 
can also see that any topological chiral edge modes 
obtained from legitimate spin-wave 
approximation appear only at a finite frequency 
region; never be a gapless mode.  

Chiral edge modes proposed in this paper share 
similar physical properties with well-known Damon-Eshbach (DE) 
surface mode.~\cite{Damon2}  
The topological modes as well as DE surface mode 
are propagating in a chiral way along 
boundaries (surfaces) of the systems, 
where the propagation   
directions are parallel (or anti-parallel)
to vector products between 
the polarized ferromagnetic moment and the normal 
vectors associated with surfaces. Experimental 
techniques for measuring DE surface mode can be also 
utilized for detecting the proposed topological chiral edge modes. 
Possible experiments include Brillouin light scattering (BLS) 
measurements,~\cite{Demokritov,Brillouin,Hillebrands}  
time-resolved Kerr microscopy,~\cite{Au}  
infrared thermography~\cite{Geisau} and scanning local magnetic 
fields in terms of a wire loop or antenna.~\cite{Vlannes} 

The topological edge mode always has a chiral dispersion 
within a band gap for volume-mode bands. When a radio 
frequency (rf) of applied microwave is chosen inside the 
band gap, spin waves are excited only along these chiral edge 
modes, while other volume modes remain intact. 
One can test this situation by changing the position of 
an input antenna from the boundaries to an interior 
far from the boundaries. Being protected by the 
topological Chern integers defined for volume modes, 
the proposed chiral edge modes are expected to be 
robust against 
various perturbations introduced near the boundaries. 
The robustness can be also tested by 
changing boundary shape or introducing boundary 
roughness and obstacle.   

Contrary to the DE mode, eq.~(\ref{general}) 
dictates that the number 
and the sense of rotation 
of the topological chiral edge modes are determined by  
the Chern integer for volume modes below the band gap. 
In fact, the chiral  edge mode in 
the proposed magnonic crystal rotates along the 
boundary in the {\it clockwise} manner with 
an up-headed magnetic field (Fig.~\ref{fig:MC}), 
while the DE surface mode with the same geometry 
rotates in the {\it counterclockwise} 
way with the up-headed field.~\cite{Damon1,Damon2} 
Moreover, the Chern integer for a volume-mode band  
itself can be changed by closing the band gap, as was 
shown in the previous section. Thus, 
using band gap manipulation, one can even control 
the chiral direction~\cite{setal} or the number 
of the mode, which enable intriguing 
spintronic device such as a spin-current 
splitter (see also sec~VI). To our best knowledge, 
the DE mode in a uniform thin film does not 
have such properties.

\section{Micromagnetic simulation for chiral topological edge modes}
\subsection{simulation procedure and material parameters}
To justify the existence of topological chiral spin-wave 
edge modes in the present MC model, 
we have numerically simulated the 
Landau-Lifshitz-Gilbert equation
\begin{eqnarray}
\frac{d{\bm m}}{dt}=-\gamma |\mu_0| 
{\bm m}\times {\bf H}_{\rm eff}+ 
\frac{\overline{\alpha}}{M_{\rm s}}{\bm m}
\times\frac{d{\bm m}}{dt}.\label{LLG}
\end{eqnarray}
where $\overline{\alpha}$ is set to 
the Gilbert damping coefficient of YIG;   
$\overline{\alpha}=6.7\times 10^{-5}$.~\cite{Kajiwara}  
The magnetic field ${\bf H}_{\rm eff}$  
includes a short-ranged 
exchange field, long-ranged dipolar field ${\bm h}$, 
the static longitudinal external field $H_0=0.1$T 
and a small temporally alternating transverse 
field $(H_1 {\bm e}_x \cos\omega t)$ with 
$H_1=1.0$ Oe. 
A simulated magnonic crystal (MC) is 
as large as either $28\mu$m $\times$ $28\mu$m 
or  $7\mu$m $\times$ $7\mu$m in 
the $x$-$y$ plane, which is composed of 
$80$ $\times$ $80$ or $20$ $\times$ $20$ 
MC unit cells respectively; each unit 
cell size is $350$ nm $\times$ $350$ nm within 
the plane. A MC unit cell consists 
of YIG region and Fe region, where the size 
of the Fe region is $140$nm $\times$ $140$ nm 
($f=0.16$). The MC studied in the previous 
sections is translationally symmetric 
along the $z$-direction. To mimic this 
situation in micromagnetic simulation, 
we take the the thickness along the $z$-direction 
$L_z$ to be sufficiently large ($L_z=1$mm). 

In actual simulation,   
each MC unit cell is further discretized into a 
bunch of smaller elements,~\cite{Dvornik} each 
of which is taken in this study as large as  
$70$ nm $\times$ $70$ nm $\times $ $L_z$ and   
each element is assigned with one ferromagnetic 
spin, which is uniformly distributed over the 
element. Namely, every MC unit cell has 25 spins 
consisting of 4 Fe spins and  21 YIG spins. 
The short-ranged exchange stiffness 
between Fe elements is taken to be 
$A_{\rm Fe}=2.1\times 10^{-11}$ J/m, 
between YIG elements $A_{\rm YIG}=0.437\times 10^{-11}$ 
J/m, and between Fe and YIG elements 
$A_{{\rm Fe-YIG}}=1.0\times 10^{-11}$ J/m.
These set material parameters to be 
those in the dipolar regime with $C_1=2$ ($H_0=0.1$T, 
$\lambda=0.35\mu $m, $r=1$ and $f = 0.16$).  

The super-elongated cell size ($70$ nm $\times$ $70$ nm 
$\times$ $L_z$ with $L_z=1$ mm) used in this simulation 
clearly ignores those spin-wave modes which have 
{\it nodes} along the 
$z$-direction. In the presence of very large thickness 
along the $z$-direction (e.g. $28\mu$m $\times$ 
$28\mu$m $\times$ $1$mm), however, 
it is likely that the strong magnetic shape anisotropy 
pushes up such spin-wave modes into higher frequency 
regimes; spin-wave excitations in lower-frequency 
regime, which we focus on 
in the present study, are mainly dominated by those 
modes having {\it no} node 
along the $z$-direction. Even for spin-wave modes 
without nodes along the $z$-direction, their 
wavefunctions should be 
certainly modified near the boundaries along the 
direction. When the thickness along the direction 
is much larger than the linear dimension within the other two 
directions ($xy$-plane), however, the modifications of the 
wavefunctions are also expected to be small. 

The time evolution is determined by Eq.~(\ref{LLG}), 
which is numerically integrated 
with a time interval of  $1\,$ps by use of 
the 4th order Runge-Kutta method. The 
demagnetization field ${\bm h}$ is calculated by the 
convolution of a kernel which describes 
the dipole-dipole interaction. 
With the Gilbert damping term, 
the system eventually reaches a certain steady 
state, in which only the spin-wave modes 
around the external frequency $\omega$ 
are excited permanently.     
In order to deduce spatial distribution of 
spin-wave modes at a given frequency, 
we have studied  
steady states ($ \omega t \gg 1$) with 
changing the external frequency.   

To compare simulation results with dispersions 
for the volume-mode bands obtained from the 
plane-wave theory, we also take a Fourier-transformation of 
transverse moments over space and time in a 
steady state. Specifically, we perform a discrete Fourier 
transformation of $m_+ \equiv m_x+im_y$ with respect to 
space and time coordinate as; 
\begin{align}
m_{+}(k_x,k_y.\omega) \equiv \sum_{X,Y} \sum_j m_{+}(X,Y,j\Delta T) 
e^{ik_x X+ik_y Y - i\omega j \Delta T}. \nn
\end{align}
A contour-plot of (absolute value of) the left hand side 
as a function of $k_x$, $k_y$ and $\omega$ is expected 
to give dispersion relations for spin-wave volume-mode 
bands.

\begin{figure}
\begin{center}
\includegraphics[scale=0.143]{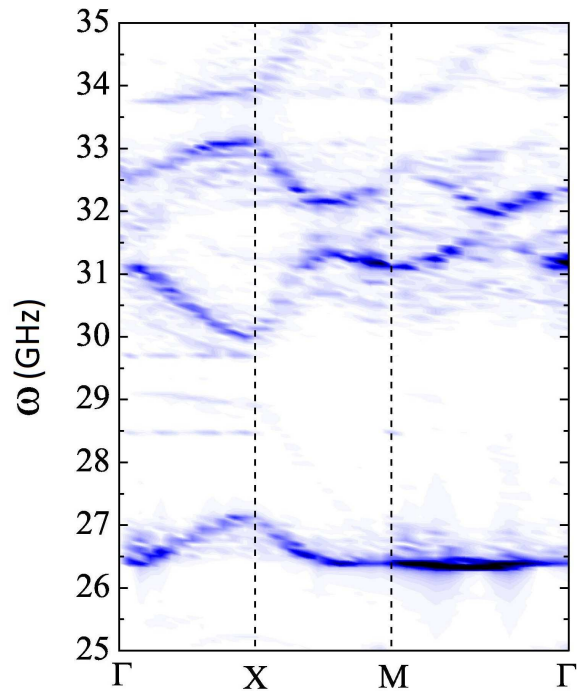}
\caption{A contour-plot of $|m_{+}(k_x,k_y,\omega)|$ 
as a function of $k_x$, $k_y$ and $\omega$, which 
has stronger amplitude in blue-colored regions 
while smaller amplitude in white regions. The contour-plot 
signifies dispersion relations for volume-mode bands (compare 
with Fig.~\ref{fig:MCband}b). The wavevector 
$(k_x,k_y)$ is along $(0,0)$ ($\Gamma$ 
point), $(\pi,0)$ ($X$ point), $(\pi,\pi)$ ($M$ point) 
and $(0,0)$ ($\Gamma$ point). In this calculation, 
the system size is taken to be 7$\mu$m$\times$ 
7$\mu$m. In this system size, we  
observe chiral edge modes similar to those in 
Fig.~\ref{fig:simulation}c,d,e,f within 
27GHz$<\omega <$30GHz.}  
\label{fig:dispersion}
\end{center}
\end{figure}   

\begin{figure}
\includegraphics[scale=0.115]{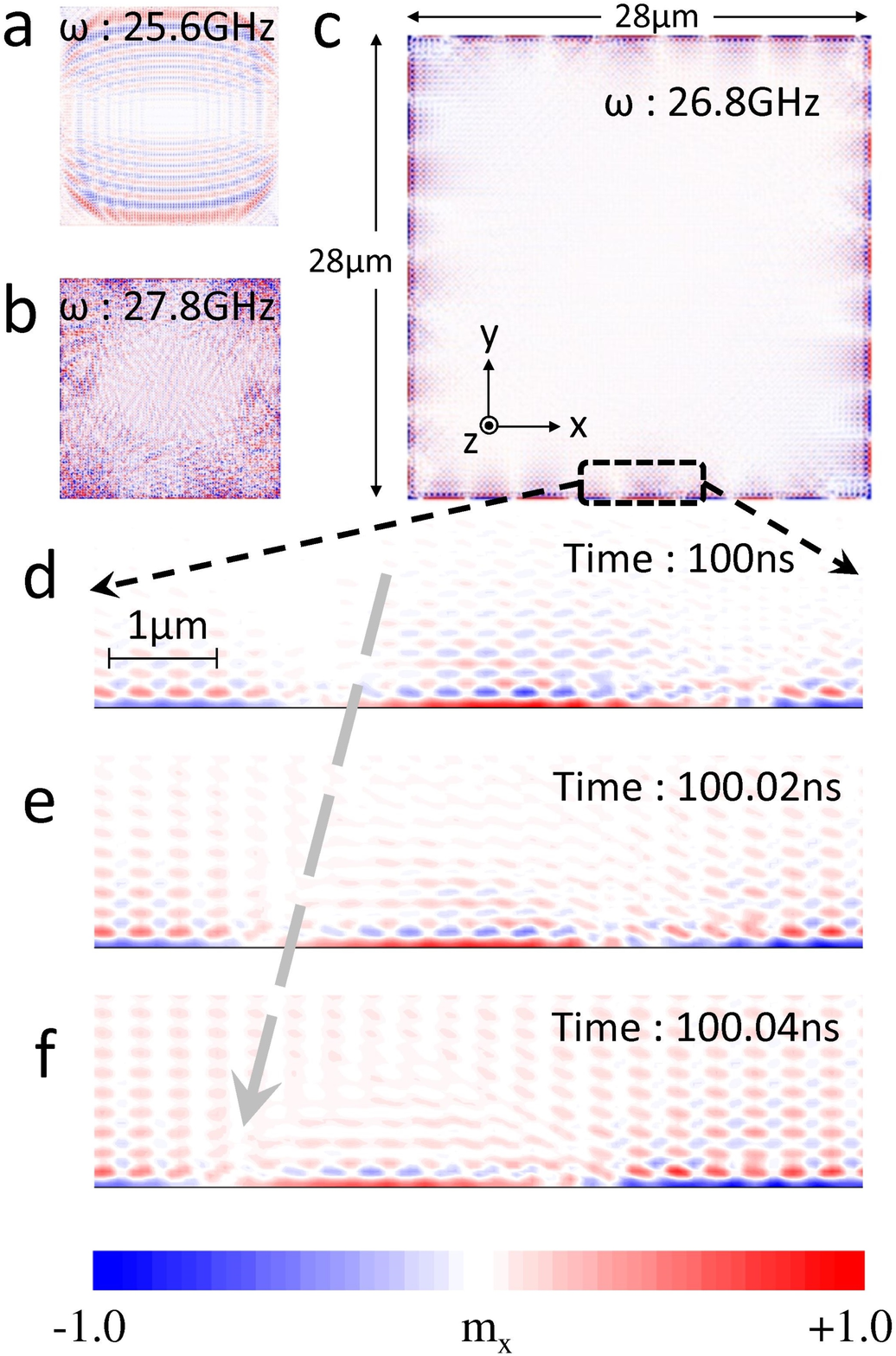}
\caption{Snapshots of spatial distribution 
of spin-wave excitations in steady states. 
In this calculation, the system size 
is taken to be 28$\mu$m$\times$ 28$\mu$m.  
A normalized transverse component of the ferromagnetic 
spin, $m_x \equiv n_x/n_{\rm max}$, are plotted, where 
$n_{\rm max}$ denotes the maximum value of 
$\sqrt{n^2_x+n^2_y}$ with $(n_x,n_y,n_z) \equiv {\bm m}/M_s$.   
The static magnetic field is 
taken along the $+z$ direction.    
\textbf{a,} snapshot of $m_x$ at $t=100\,$ns 
with the external frequency  $\omega_0<\omega<\omega_1$
($\omega=25.6\,$GHz). 
{\textbf b,} snapshot at 
$t=100\,$ns with 
$\omega_2<\omega$ ($\omega=27.8\,$GHz). 
{\textbf c}, snapshot at $t=100\,$ns 
with $\omega_1<\omega<\omega_2$ 
($\omega=26.8\,$GHz).  \textbf{d,e,f,} 
Spin-wave edge modes at $\omega=26.8$ GHz 
are propagating in a chiral way (\textbf{d,} 
$t=100$ ns, \textbf{e,} $t=100.02$ ns. 
\textbf{f,} $t=100.04$ ns)}  
\label{fig:simulation}
\end{figure}

\subsection{result}
Throughout the micromagnetic simulation, 
we found that a lower frequency region can be 
roughly classified into three 
characteristic regimes; two volume-mode 
frequency regimes, (i) $ \omega_0 < \omega < \omega_1$ 
and (iii) $ \omega_2 < \omega$, and    
 a band-gap regime for volume modes, 
(ii) $\omega_1 < \omega < \omega_2$.
Below  $\omega_0$, 
the system remains intact against  the 
small alternating transverse field, indicating no 
spin-wave modes for $\omega<\omega_0$. 
Within the volume-mode 
frequency regimes (i) $ \omega_0 < \omega < \omega_1$ (Fig.~\ref{fig:simulation}a)
and (iii) $\omega_2 < \omega$ (Fig.~\ref{fig:simulation}b), 
spin-wave excitations in steady states 
are always distributed over the entire system. 
A comparison between Fig.~\ref{fig:MCband}b  
and the contour-plot of the Fourier-transform of the 
transverse moments (Fig.~\ref{fig:dispersion}) 
indicates that these two frequency regimes 
correspond to the lowest volume-mode band and 
higher volume-mode bands respectively.   

On the one hand, a steady state in the intermediate 
frequency regime, $\omega_1 < \omega < \omega_2$, 
has almost no weight for volume modes.  
Instead, spin-wave excitations in the 
intermediate regime are localized only around the boundaries 
of the system (see Fig.~\ref{fig:simulation}c), indicating the 
existence of spin-wave edge modes. Moreover, 
these edge modes propagate in a {\it chiral} way 
(see Fig.~\ref{fig:simulation}d,e,f), whose 
direction is consistent with a chiral direction  
determined from the Chern 
integer in the dipolar regime found in 
Fig.~\ref{fig:MCband}a, $C_1=+2>0$. 
These observations indicate that 
the spin-wave edge modes found in the intermediate 
frequency region is nothing but the topological chiral 
spin-wave edge modes described in the previous section.
In fact, the three frequency regimes are qualitatively 
consistent with the plane-wave-theory calculation 
(Fig.~\ref{fig:MCband}b); $(\omega_0,\omega_1,\omega_2) = 
(25.5 {\rm GHz},26.0{\rm Ghz}, 27.5{\rm GHz})$ 
for $28\mu$m $\times 28\mu$m $\times 1$mm, 
and (26.2 {\rm GHz},27.1{\rm GHz}, 30.0{\rm GHz})  
for $7\mu$m $\times 7\mu$m $\times 1$mm. 
Our micromagnetic simulation  
with a  shorter sample thickness ($L_z=200\mu$m) 
also justifies the existence of topological chiral modes.

\section{Discussion}  
By calculating a newly-introduced bosonic Chern 
integer for spin wave bands, we argue that 
a two-dimensional normally-magnetized 
magnonic crystal acquires chiral  
edge modes for magnetostatic wave in the 
dipolar regime. Each mode is localized 
at the boundary of the system, carrying magnetic 
energies in a unidirectional way. 
Thanks to the topological protection, spin-wave 
propagations along these edge modes are  
robust against imperfections of the 
lattice periodicity and boundary roughness; 
they are free from any types of elastic backward 
scatterings with moderate strength.~\cite{Halperin} 
This robustness makes it possible to implement 
novel fault-tolerant magnonic device such as spin-wave 
current splitter and a magnonic 
Fabry-Perot interferometer as discussed below.

\begin{figure}
\includegraphics[scale=0.16]{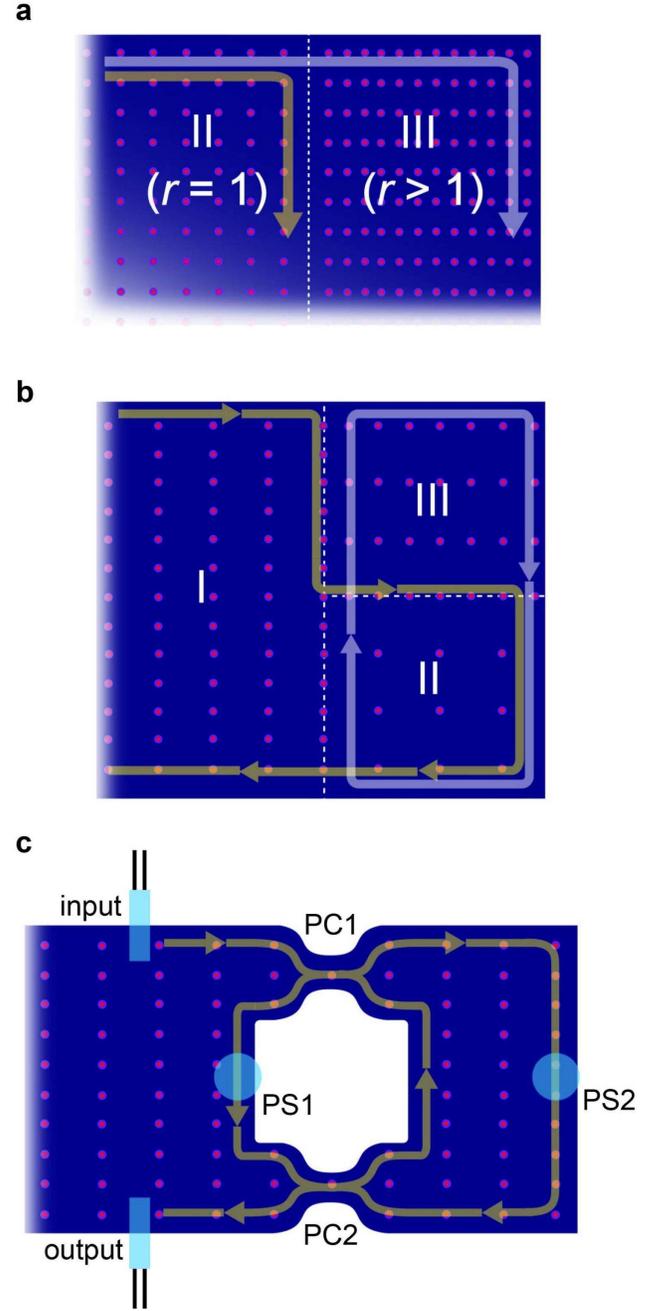}
\caption{Examples of magnonic  
circuit made by chiral magnonic edge modes. 
\textbf{a,b,} schematic pictures of spin-current splitter. 
{\textbf c}, Magnonic analogue 
of the Fabry-Perot interferometer.}
\label{fig:splitter}
\end{figure}

The chiral spin-wave edge modes studied in this paper 
can be easily twisted or split by changes of the size
($\lambda$) and shape ($r$) of the unit cell,  
which we demonstrate in Fig.~\ref{fig:splitter}a,b. 
In Fig.~\ref{fig:splitter}a, 
the MC in the phase II ($r=1$) is connected 
with the other MC in the phase III ($r>1$), whose 
Chern integer for the lowest band differ by unit.   
A boundary between these two MC systems supports 
the chiral spin-wave edge mode which runs across 
the direct band gap at $(0,\pi)$ point. 
This means that the two chiral edge modes 
propagating along the boundary of the 
the MC in the phase II ($r=1$) are spatially divided 
into two, where one mode goes along the  
boundary of the MC in the phase III ($r>1$), 
while the other goes along the boundary between 
these two MCs. This configuration realizes a 
spin-wave current splitter, an alternative to those 
proposed in other geometries.~\cite{Demidov}

The Fabry-Perot interferometer is made up 
of a coupled of chiral spin-wave edge 
modes encompassing a single topological MC 
(see Fig.~\ref{fig:splitter}c). 
Parts of the MC are spatially 
constricted by the hole inside, so as to play  
a role of the point contact between 
these edge modes.~\cite{Ji} A unidirectional 
spin-wave propagation is induced in a chiral 
mode via either an antenna 
attached to the boundary (`input' in Fig.~\ref{fig:splitter}c)  
or a microwave-spin-wave transducer put 
near the boundary.~\cite{Au,Au2} 
The wave is divided into two 
chiral edge modes at a point contact 
(`PC1' in Fig.~\ref{fig:splitter}c). 
Two chiral propagations merge 
into a single chiral propagation at the 
other point contact (`PC2'). 
Depending on a phase 
difference between these two, 
the superposed wave exhibits 
either a destructive or a constructive 
interference, which is detected as an 
electric signal from the other antenna 
(`output'). Note that 
local application of magnetic fields   
(`PS1' or `PS2' in Fig.~\ref{fig:splitter}c) change  
 the velocities of the two chiral 
edge modes locally. These modifications result 
in phase shifts of the two chiral waves, which thus 
changes the interference pattern observed in the 
output signal. With the use of these local magnetic 
fields as an external control,~\cite{Schneider,Kruglyak} 
the interferometer can serve as 
a solid-state based magnonic logic gate. 
A combination with a recently-proposed 
resonant microwave-to-spin-wave 
transducer~\cite{Au2} would also expand further 
prospects of spintronic applications of these 
spin-wave devices. 

Though chiral edge modes are robust against 
static perturbations, 
magnetic energies excited in the edge mode 
decay into either phonon states or other magnon 
states in the volume modes via inelastic scatterings. The 
associated decay time or coherence 
length depends on specific materials, and spin wave 
propagation along the chiral edge mode 
survives only within this coherence length. 
Due to the absence of conduction electrons, however, 
spin waves in magnetic insulators 
have long coherence lengths; The coherence length in YIG 
in the magnetostatic regime becomes 
on the order of centimeter.~\cite{Serga}  
In such magnetic insulators, the characteristic 
spin-wave propagations depicted 
in Fig.~\ref{fig:splitter}a,b,c are experimentally 
realizable especially in sizable MC systems. 
Experimental measurement of these propagations 
in the space- and time-resolved manner 
is by itself remarkable, which surely 
leads to the development of 
innovative spintronic device 
in future.  

After submission of the present manuscript into the 
preprint server,~\cite{SMM} 
we found another submission,~\cite{DM} 
which also theoretically explored the 
realization of similar topological 
chiral magnonic edge mode in localized spin 
system. Their edge modes come from the 
short-ranged Dzyaloshinskii-Moriya 
exchange interaction 
instead of the (more classical) magnetic 
dipole-dipole interaction studied here.

\begin{acknowledgements}
We would like to thank Masayuki Hashisaka for 
discussions. RS also thanks to Tsutomu Momoi 
for informing him of Ref.~\onlinecite{Colpa}. 
This work is supported in part 
by Grant-in-Aids  
from the Ministry of Education,
Culture, Sports, Science and Technology of Japan
 (No.~21000004, 22540327, 23740284 and 
24740225) and by 
Grant for Basic Scientific Research Projects 
from the Sumitomo Foundation.  
\end{acknowledgements}

\appendix
\section{Magnetic Monopole and Dirac Hamiltonian}
When two bosonic (magnonic) bands form a 
band-touching point in the 
3-dimensional ${\bm p}$-parameter space 
with ${\bm p} \equiv (k_x,k_y,\lambda)$,  
the dual magnetic fields for these two bands 
(Eqs~(\ref{Ch4},\ref{25},\ref{gauge}))  
have a quantized source of their 
divergence at the point 
(${\bm p}={\bm p}_c$). Away from the band-touching 
point, the dual gauge fields (Eqs~(\ref{Ch4},\ref{gauge})) 
can be locally determined,  so that the 
dual magnetic fields (Eq.~(\ref{25})) are clearly 
divergence-free. At ${\bm p} = {\bm p}_{c}$ 
the projection to each of these two bands  
cannot be defined, which 
endows the respective dual magnetic field 
with some singular structure.  
The singular structure can be studied by  
the degenerate perturbation theory 
for a generalized eigenvalue problem. 
The eigenvalue problem takes a form  
\begin{eqnarray}
{\bm H}_{\bm p} {\bm T}_{\bm p} = 
{\bm \sigma}_3 {\bm T}_{\bm p}  
\left[\begin{array}{cc} 
{\bm E}_{{\bm p}} & \\
& - {\bm E}_{\overline{\bm p}} \\
\end{array}\right], \nn 
\end{eqnarray}
with ${\bm p}\equiv (k_x,k_y,\lambda)$ 
and $\overline{\bm p} \equiv (-k_x,-k_y,\lambda)$.  
The diagonal matrix ${\bm \sigma}_3$ takes 
$+1$ for the particle space while 
takes $-1$ in the hole space. 
${\bm H}_{\bm p}$ is a quadratic form 
of boson Hamiltonian introduced in sec. I. 
${\bm E}_{{\bm p}}$ is a diagonal matrix, 
whose elements give dispersions for 
bosonic (magnonic) bands and are physically 
all positive definite. We decompose this into 
the zero-th order part and the perturbation part; 
\begin{eqnarray}
{\bm H}_{{\bm p}} = {\bm H}_{0} + 
{\bm V}_{{\bm p}}. \label{n}
\end{eqnarray} 
with ${\bm H}_{0}\equiv {\bm H}_{{\bm p}={\bm p}_c}$ 
and ${\bm V}_{\bm p}\equiv {\bm H}_{\bm p}-{\bm H}_0$.  
Suppose that ${\bm H}_{0}$ has 
two-fold degenerate eigenstates ${\bm t}_j$ 
($j=1,2$) 
with its eigen-frequency $\omega_{0}(>0)$;  
\begin{eqnarray}
{\bm H}_{0} \!\ {\bm t}_j   
= {\bm \sigma}_3 \!\ {\bm t}_{j}   
\!\  \omega_{0}, \nn
\end{eqnarray} 
where the states are normalized as  
${\bm t}^{\dagger}_{j}{\bm \sigma}_3 
{\bm t}_{m} = \delta_{jm}$. 
On introducing the perturbation ${\bm V}_{\bm p}$, 
the degeneracy is split into two frequency levels. 
The eigenstate for the respective 
eigen-frequency is determined on the 
zero-th order of ${\bm p}-{\bm p}_c$ as;
\begin{eqnarray}
{\bm T}_{\bm p} = {\bm T}_{0} {\bm U}_{\bm p} 
+ {\cal O}(|{\bm p}-{\bm p}_c|), \label{appr} 
\end{eqnarray} 
where ${\bm T}_0$ diagonalizes 
${\bm H}_0$ with 
${\bm T}^{\dag}_{0}{\bm \sigma}_3 {\bm T}_{0} 
= {\bm T}_{0}{\bm \sigma}_3 {\bm T}^{\dag}_{0}
= {\bm \sigma}_3$ and a unitary matrix ${\bm U}_{\bm p}$ 
diagonalizes a 2 by 2 Hamiltonian 
${\bm V}_{\rm eff}$ formed by the 
two-fold degenerate eigenstates; 
\begin{eqnarray}
{\bm V}_{\rm eff} \equiv \left[\begin{array}{cc} 
{\bm t}^{\dagger}_{1}{\bm V}_{\bm p} 
{\bm t}_{1} & 
{\bm t}^{\dagger}_{1}{\bm V}_{\bm p} 
{\bm t}_{2} \\
{\bm t}_{2}{\bm V}_{\bm p} 
{\bm t}_{1} & 
{\bm t}^{\dagger}_{2}{\bm V}_{\bm p} 
{\bm t}_{2} \\
\end{array}\right]. \label{Veff}
\end{eqnarray}
By substituting Eq.~(\ref{appr}) into 
Eqs.~(\ref{Ch4},\ref{25},\ref{gauge}), 
one can easily see that, near ${\bm p}={\bm p}_c$, 
the dual magnetic field is 
given only by the unitary matrix; in the leading 
order of ${\bm p}-{\bm p}_c$, it is given as 
\begin{align}
{\bm B}_{j} = {\bm \nabla} \times {\bm A}_{j} + 
{\cal O}(|{\bm p}-{\bm p}_c|^{-1}), \ \ 
{\bm A}_{j} = i {\rm Tr}[{\bm \Gamma}_{j} 
{\bm U}^{\dag}_{\bm p}{\bm \nabla}{\bm U}_{\bm p}], \nn 
\end{align} 
with ${\bm \nabla} \equiv 
(\partial_{k_x},\partial_{k_y},\partial_{\lambda})$. 
Now that Eq.~(\ref{Veff}) reduces to 
a usual 2 by 2 Dirac-type Hamiltonian, we 
can show the quantization of the dual 
magnetic charge at the band-touching 
point exactly in the same way as 
in the 2 by 2 Dirac fermion system.~\cite{Berry,Simon} 
Thereby, the sign and the strength of the magnetic 
charge is determined only by the $2$ by $2$ effective 
Dirac Hamiltonian. With a proper gauge 
transformation and scale transformation, 
the effective Hamiltonian at the band-touching 
points $P_{j}$ ($j=1,2$) 
takes a form;
\begin{align}
{\cal H}_{\rm eff} = \omega_0 {\bm \tau}_0 
+ (\lambda-\lambda_{c,j}) {\bm \tau}_3 
+ a p_x {\bm \tau}_1 + b p_y {\bm \tau}_2, \nn
\end{align}
with $a>0$, $b>0$, 
$(p_x,p_y) \equiv (k_x-\pi,k_y)$ for $j=1$ 
and $(p_x,p_y) \equiv (k_x,k_y-\pi)$ for $j=2$.  
Eq.~(\ref{effective1}) is derived by the replacement of 
$p_\mu \rightarrow -i\partial_{\mu}$.

\
 
\ 

\noindent
\end{document}